\documentclass[reprint,amsmath,amssymb,aps]{revtex4-2}
\newcommand{\be}{\begin{equation} } 
\newcommand{\ee}{\end{equation} } 
\newcommand{\ba}{\begin{array} } 
\newcommand{\ea}{\end{array} } 
\newcommand{\bear}{\begin{eqnarray} } 
\newcommand{\eear}{\end{eqnarray} } 
 
\newcommand{\ms}{M_S}
\newcommand{\sev}{S_{\rm ev}}
\newcommand{\bev}{B_{\rm ev}}

\usepackage{graphicx, comment, multirow, appendix, tikz-feynman, tikz}
\usepackage{dcolumn}
\usepackage{bm}
\usepackage{hyperref}
\usepackage{lineno}
\usepackage{float}

\begin{document}

\vspace*{-1.cm}

\noindent \makebox[5.cm][l]{\small \hspace*{-.2cm} }{\small Fermilab-Pub-25-0149-T}  \\  [-1mm]

\title{Ultraheavy diquark decaying into vectorlike quarks at the LHC}
\author{Ioana Duminica,$^{1,2}$ Calin Alexa,$^1$ Ioan M. Dinu,$^1$ Bogdan A. Dobrescu,$^3$ and Matei-Stefan Filip,$^{1,2}$}
\email{Corresponding author: calin.alexa@cern.ch}
\affiliation{
$^1$Particle Physics Department, \href{https://ror.org/00d3pnh21}{IFIN-HH},  M\u agurele, IF 077125, Romania \\
$^2$Faculty of Physics, \href{https://ror.org/02x2v6p15}{University of Bucharest}, M\u agurele, IF 077125, Romania \\
$^3$Particle Theory Department, \href{https://ror.org/020hgte69}{Fermilab}, Batavia, Illinois 60510, USA  
}
\date{March 21, 2025; Revised August 12, 2025}

\begin{abstract}
We explore the discovery potential of an ultraheavy $(7-8.5$ TeV) diquark scalar
produced in the collisions of two up quarks at the LHC.
Assuming that the diquark scalar decays into two vectorlike quarks of mass around 2 TeV, each of them decaying into a $W^{+}$ boson and a $b$ quark, we focus on the fully-hadronic final state. We present a signal-from-background separation study based on a discriminator built with Machine Learning techniques. For this six-jet final state and a luminosity of $3000 \ \text{fb}^{-1}$, we estimate that a diquark scalar of mass near 8 TeV may be discovered or ruled out even when its coupling to up quarks is as low as 0.2.\\

DOI: \href{https://journals.aps.org/prd/accepted/10.1103/hxll-41ly}{10.1103/hxll-41ly}
\end{abstract}

\maketitle

\section{\label{sec:intro}Introduction}
Starting with the ongoing Run 3 data taking at the LHC and the upcoming high-luminosity LHC (HL-LHC)~\cite{ZurbanoFernandez:2020cco}, explorations of physics near the 10 TeV scale become possible \cite{Dobrescu:2019nys},
and the discovery potential for physics beyond the Standard Model (SM) considerably increases, especially for ultraheavy particles near the kinematic limit of the collider.

In Run 2 of the LHC, the CMS Collaboration \cite{CMS:2022usq} observed two events with four hadronic jets characterized by very large tetrajet invariant mass $M_{4j}$, around 8 TeV. 
Moreover, in both events, the four jets are paired in two dijets, each having an invariant mass $M_{2j}$ in the $1.9 - 2.1$ TeV range. The QCD background for such events is of the order of $10^{-4}$ events \cite{Dobrescu:2018psr}, and the CMS Collaboration concluded that the two observed events constitute a $3.9\sigma$ excess over the SM prediction. 
Furthermore, CMS interpreted the excess in terms of a model presented in \cite{Dobrescu:2018psr}, in which a diquark scalar $S_{uu}$ is produced in the $s$-channel from a $uu$ initial state, and decays into a pair of vectorlike quarks $\chi$. The subsequent decay $\chi \to u g$, which occurs at one loop \cite{Dobrescu:2024mdl}, leads to events consistent with the observed $4j$ events provided $S_{uu}$ and $\chi$ have masses $\ms\approx 8.5$ TeV and $m_\chi\approx 2.1$ TeV, respectively.

A study similar to the CMS $4j$ search was performed by the ATLAS Collaboration \cite{ATLAS:2023ssk}. 
An outlier 4-jet event with $M_{4j} \approx 6.6$ TeV was observed in that study, consisting of two dijets of $M_{2j} \approx 2.2$ TeV each. Although the 4-jet invariant mass of the ATLAS event is significantly below the ones of the CMS events, it is still consistent (roughly at the 1.5$\sigma$ level) with the $u u \to S_{uu} \to \chi\chi \to 4j$ hypothesis employed by CMS, because the $4j$ invariant mass distribution produced by an ultraheavy $S_{uu}$ has a long tail toward lower masses.

In this paper we study the LHC reach in a different possible final state of an ultraheavy $S_{uu}$, which arises from a more typical decay mode of the vectorlike quark $\chi$, namely $S_{uu} \rightarrow\chi\chi\rightarrow (W^+b)(W^+b)$. The $\chi \to W^+b$ decay is unavoidable if there is mass mixing between $\chi$ and the SM top quark. An alternative origin of the $\chi \to W^+b$ decay may be a loop process, similar to the one responsible for $\chi \to u g$.

In either case, additional decay modes of $\chi$ are expected (such as $Z t $ and $h^0 t $; the sum of these two branching fractions is approximately equal to the $W^+b$ one in the case of mass mixing \cite{Han:2003wu}), but they are more model dependent, and will be not be studied here.

We will focus mainly on hadronic final states of the $W$ bosons arising from the $\chi$ decays, because these allow the reconstruction of the $S_{uu}$ mass.
Signal search optimization is done with Machine Learning (ML) methods, 
a powerful approach in particle and event identification and reconstruction
largely used in multivariable environments  \cite{Vidal:2021oed, Albertsson:2018maf}.

In Sec. \ref{sec:particles} we describe the interactions of the new particles, and the simulation of the signal and backgrounds.
The ML discriminant is discussed in Sec. \ref{sec:ML}, the selection efficiency is estimated in Sec. \ref{sec:effic}, and
the $S_{uu}$ observation potential is investigated in Sec. \ref{sec:obs_potential}.
Our conclusions are presented in Sec. \ref{sec:conclusions}.

\bigskip 

\section{A diquark and a vectorlike quark}
\label{sec:particles}

The model considered here \cite{Dobrescu:2019nys} (see also \cite{Dobrescu:2018psr}) includes two particles beyond the SM: a diquark scalar, $S_{uu}$, and a vectorlike quark, $\chi$.
The $S_{uu}$ diquark is a color-sextet weak-singlet complex scalar particle that carries electric charge $4/3$.
The $\chi$ vectorlike quark is a color-triplet weak-singlet fermion that carries electric charge $2/3$.

The diquark has interactions with two up quarks, and also with two $\chi$ quarks, described by the following terms in the Lagrangian:
\be
\dfrac{y_{uu}}{2} \,  S_{uu} 
\overline u_{R} \,  u^c_{R} \, + 
\dfrac{y_{\chi\chi}}{2} \,  S_{uu}  \overline \chi_{R} \,  \chi^c_{R} \, + {\rm H.c.}
\label{eq:diquarkYuk}
\ee
Here the Yukawa couplings $y_{uu}$ and $y_{\chi\chi}$ are parameters of order one, or smaller. 
The upper index $c$ refers to charge conjugation, so that the Feynman rule associated with a Lagrangian term of the type $S_{uu} 
\overline \chi_{R} \,  \chi^c_{R}$ is such that one scalar enters the interaction vertex and two quarks exit from the vertex.

Other possible Yukawa interactions of $S_{uu}$, discussed in \cite{Dobrescu:2019nys}, are assumed here to have negligible couplings.
The $\chi$ interaction with the $W$ boson is taken to be 
\be
g_\chi \, W^+ \bar\chi_L \gamma^\mu b_L
+ {\rm H.c.} ~~,
\label{eq:chiW}
\ee
where $g_\chi \ll 1$. This may arise, for example, from mass mixing between $\chi$ and the SM top quark \cite{Dobrescu:2009vz}. Additional interactions of a single $\chi$ with a SM quark and a SM boson are expected, but their effects are taken into account here only indirectly, through a branching fraction $B(\chi \to W b) < 1$. 

\begin{figure}
    \centering
\begin{tikzpicture}
  \begin{feynman}
    \vertex (z1);
    \vertex [right=1.5cm of z1] (z2);
    \vertex [above left=1.0cm and 1.5cm of z1] (e1) {\(u\)};
    \vertex [below left=1.0cm and 1.5cm of z1] (e2) {\(u\)};

    \vertex [above right=1.0cm and 1.5cm of z2] (x1);
    \vertex [below right=1.0cm and 1.5cm of z2] (x2);

    \vertex [below right=0.25cm and 1cm of x1] (x3) {\(W^{+}\)};
    \vertex [above right=0.25cm and 1cm of x1] (q12) {\(b\)};
    \vertex [above right=0.25cm and 1cm of x2] (x4) {\(W^{+}\)};
    \vertex [below right=0.25cm and 1cm of x2] (q22) {\(b\)};
    \diagram* {
      (e1) -- [fermion] (z1),
      (e2) -- [fermion] (z1),

      (z1) -- [scalar, edge label=\(S_{uu}\)] (z2),
    
      (z2) -- [fermion, edge label=\(\chi\)] (x1),
      (z2) -- [fermion, edge label'=\(\chi\)] (x2),

      (x1) -- [boson] (x3),
      (x2) -- [boson] (x4),

      (x1) -- [fermion] (q12),
      (x2) -- [fermion] (q22),
    };
  \end{feynman}
\end{tikzpicture}
    \caption{Diquark $S_{uu}$ production at the LHC, followed by a cascade decay through two vectorlike quarks. The final state is $W^+ b \, W^+ b$.}
    \label{fig:diagram}
\end{figure}
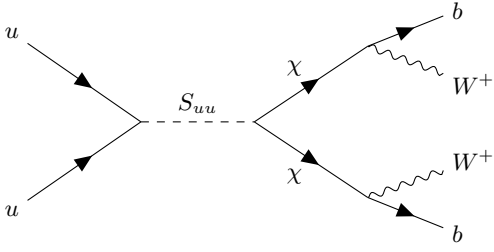

The main production mechanism for $S_{uu}$ at proton-proton colliders is from a $uu$ initial state. Its subsequent decay into $\chi\chi$, followed by each $\chi$ decaying into $W^+b$, is represented by a single diagram, shown in Fig.~\ref{fig:diagram}.
The cross section for this 
$pp\to S_{uu}\to \chi\chi \to (W^+b) \, (W^+b)$ process, denoted here as $\sigma_{\! _S}(Wb \, Wb)$, is given by 
\be
\sigma_{\! _S\,}\!(Wb \, Wb) \approx
\sigma_{\! _S\,}\!(\chi\chi)
\, B(\chi\to Wb)^2  ~~,
\label{eq:rate}
\ee
where $\sigma_{\! _S}(\chi\chi) \equiv \sigma (pp\to  S_{uu} \! \to\chi\chi)$   is the production cross section for 
two $\chi$ quarks (not for $\chi\overline \chi$), which proceeds through an $s$-channel $S_{uu}$.
Corrections due to similar processes with an off-shell 
$\chi$ are neglected here because the vectorlike quark is expected to have a very small decay width (but still sizable enough to allow prompt decays).

Four parameters control the characteristics of the $u u \to S_{uu} \!\to \chi \chi 
$ process: the $\ms$ and $m_\chi$ masses, and the $y_{uu}$ and $y_{\chi\chi}$ couplings introduced in Eq.~(\ref{eq:diquarkYuk}).
The branching fraction of the diquark into vectorlike quarks, $B(S_{uu}\! \to \chi\chi)$, is a function of the coupling ratio $y_{\chi\chi}/y_{uu}$  and of the mass ratio $m_\chi/\ms$, and can be straightforwardly computed using the formulas for the $S_{uu}\! \to \chi\chi$ and $S_{uu}\! \to uu$ widths given in \cite{Dobrescu:2024mdl}.
For example, when the  ratios are $m_\chi/\ms = 1/4$ and $y_{\chi\chi}/y_{uu} = 1.5$, the branching fraction is $B(S_{uu}\! \to \chi\chi) \approx 63\%$.

The $\sigma_{\! _S}(\chi\chi)$ cross section is approximately given by the product of the on-shell $S_{uu}$ production cross section (computed in \cite{Dobrescu:2018psr} based on the NLO formulas derived in \cite{Han:2009ya})
times the  branching fraction of the diquark: $\sigma_{\! _S}(pp \to S_{uu})\, B(S_{uu}\! \to \chi\chi)$. 
There are non-negligible effects due to off-shell $S_{uu}$ exchange, but these are relevant mostly for  events with lower-energy final-state particles, where the backgrounds are larger.   
Note that $\sigma_{\! _S}(pp \to S_{uu})$ is fast falling with increased $\ms$, and is proportional to $y_{uu}^2$. Thus, from
Eq.~(\ref{eq:rate}) it follows that $\sigma_{\! _S}(Wb \, Wb)$ is approximately proportional to the square of  $y_{uu}B(\chi \to W b)$.
 
For ML discriminant (Sec. \ref{sec:ML}) and signal selection efficiency studies (Sec. \ref{sec:effic}),
we vary $\ms$ in the $7-8.5$ TeV range, and fix the $\chi$ mass and the coupling ratio  at 
\be
m_\chi = 2 \; {\rm TeV},  \;\;\;  \frac{y_{\chi\chi}}{y_{uu}}  = 1.5. 
\label{eq:benchmark}
\ee
Note that the CMS \cite{CMS:2022usq} and ATLAS \cite{ATLAS:2023ssk} events with dijet pairs discussed above indicate a dijet mass of 2 TeV, with uncertainties below  10\%. By contrast, the $4j$ masses of those events are consistent with a larger range of $M_S$ values, justifying the above mass choices.

If $y_{uu}$ were larger than $y_{\chi\chi}$, then $S_{uu}$ would likely be first discovered through dijet resonance searches, which is not the focus of this study. In the other limit, $y_{\chi\chi}\gg y_{uu}$, the diquark becomes a broad resonance \cite{Dobrescu:2024mdl} (since diquark production becomes too small for $y_{uu}$ below 0.2 or so), which again is not our focus. Thus,  the value of the coupling ratio specified in (\ref{eq:benchmark}) is natural. Moreover, its exact value does not significantly affect the signal discussed here, because $y_{\chi\chi}/y_{uu}$ only determines the diquark branching fractions; a change in them can be compensated by a change in the diquark production cross section (controlled by $y_{uu}$).    
We consider two values for the product of the $S_{uu}$ coupling to up quarks with the branching fraction of $\chi$:
\be
y_{uu} B(\chi \to W b) = 0.1 \, ,\; 0.2 ~.
\label{eq:benchmark_sensitivity}
\ee
Hence, for the typical value $B(\chi \to W b) \approx 50\%$ \cite{Han:2003wu,Dobrescu:2009vz}, the benchmark values for the coupling that we will consider here are $y_{uu} = 0.2$ and 0.4.
While values of $y_{uu}$ as large as 0.9 are still viable \cite{Dobrescu:2024mdl}, they will be probed soon in Run 3 searches for dijet resonances, and will not be considered here.

The model described above, including the scalar diquark plus the vectorlike quark, was implemented in FeynRules \cite{Alloul:2013bka}, which outputs the \textsc{MadGraph} model files. For signal simulation we used  \textsc{MadGraph5\_aMC@NLO} \cite{Alwall:2014hca} (v3.3.2) at leading order (LO) with the default PDF set (NNPDF23LO \cite{NNPDF:2014otw}), and the output interfaced to \textsc{Pythia8.310} \cite{Bierlich:2022pfr}. Detector response was simulated using \textsc{Delphes 3.5.0} \cite{deFavereau:2013fsa} with both ATLAS and CMS parametrizations. 
Jets were reconstructed with \textsc{FastJet} \cite{Cacciari:2011ma} using the anti-k$_{t}$ algorithm \cite{Cacciari:2008gp}. 

\begin{figure}[ht]
\centering
\vspace*{3mm}
\hspace*{-3mm}
\includegraphics[width=0.43\textwidth]{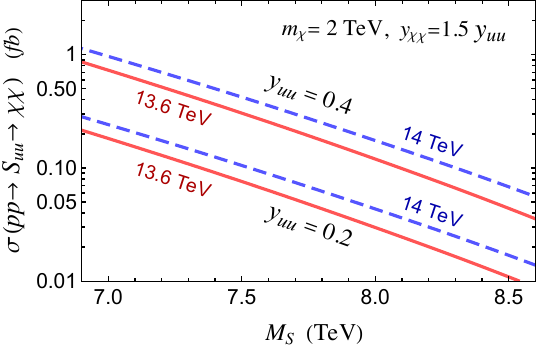} 
\vspace*{-3mm}
\caption{Cross section for the 
$pp\to S_{uu}\to \chi\chi$ process as a function of the $S_{uu}$ diquark mass, at center-of-mass energies of 13.6 TeV (solid red lines) and 14 TeV (dashed blue lines). 
This $\sigma_{\! _S}(\chi\chi)$
is computed at LO with \textsc{MadGraph} \cite{Alwall:2014hca}  
for two values of the $S_{uu}$ coupling to up quarks: 
$y_{uu} =0.2$ (bottom two lines) and 0.4 (top two lines).
Other parameters are fixed as in Eq.~(\ref{eq:benchmark}). }
\label{fig:cross_section_yuuB}
\end{figure}

In Fig. \ref{fig:cross_section_yuuB} we show the $\sigma_{\! _S}(\chi\chi)$ cross section computed with \textsc{MadGraph} as a function of $\ms$ for $y_{uu} = 0.2$ (lower lines) and $y_{uu} = 0.4$ (upper lines); the values for other model parameters are fixed there as in Eq.~(\ref{eq:benchmark}). 
Signal and background data samples were generated at $\sqrt s=13.6$ and $14$ TeV without pileup.

For background simulation we use the following types of processes: 
$2\rightarrow 2$ QCD, $W+$jets, Higgs processes, dibosons, and $t\bar{t}$; altogether there are 32 background  processes  \cite{Git:2023V2}.
For the signal selection relevant here, the most important background  processes are $2\rightarrow 2$ QCD, $W+$jets and dibosons.
Due to the very high $\ms$ values, 
we constrain the phase space for background simulation by imposing high values for the \textsc{Pythia} cut $mHatMin$ \cite{Bierlich:2022pfr} (which  represents the minimum invariant mass of the final state):   
$\widehat{m}_{\rm min}\in\left[5.5,8\right]$ TeV.

The signal considered here consists in two pairs of nearly collinear quark jets (originating from the two $W^+$ bosons) plus two $b$ jets. Thus, besides the $2\rightarrow 2$ QCD background included in our study, which is due to the showering of four extra jets, there are a few other QCD backgrounds that mimic our signal. These include the $2\rightarrow k_j$ QCD processes with $k_j = 3,4,5$ and the showering of extra jets, as well as the $2\rightarrow 6$ QCD processes. 
These additional QCD backgrounds, which we ignore, are difficult to simulate given the very large $p_T$ associated with the signal produced by an ultraheavy $S_{uu}$; nevertheless, we do not expect that they are substantially larger than the background due to $2\rightarrow 2$ QCD plus showering, which is included simulated with \textsc{Pythia}.

\section{\label{ML}Machine learning discriminant}\label{sec:ML}

Due to the complexity of the background relevant to this  study, we decided to use Machine Learning algorithms to discriminate between signal and background rather than a cut-based analysis. We tested two ML classification models~\cite{Git:2023}: Boosted Decision Tree (BDT)~\cite{Friedman:2001wbq} and Random Forest (RF)~\cite{Ho:1995} (neural networks were also tested in a preliminary attempt \cite {Duminica:2024nos}).

The BDT approach is the most commonly used for such classification tasks, relying on sequentially training multiple decision trees, where each new tree aims to correct the errors of the previous ones. Due to its fast training time, we choose the \textsc{XGBoost} \cite{Chen:2016:XST:2939672.2939785} implementation of the BDT. 
In the case of the RF models, which also consist of ensembles of decision trees, the training is performed on different subsets of data and features. Their outputs are combined  to make more accurate and robust predictions. Here, we use the \textsc{Scikit-learn}~\cite{Pedregosa:2011} implementation of the RF.

To obtain the ML discriminator, $D$, we input a selection of 75 variables into ML models:
jet kinematics $p_\text{T}^{(i)}$, $\eta^{(i)}$, $\phi^{(i)}$, 
dijet angular distance $\Delta R^{(i,j)}$, 
dijet invariant mass $m_{2j}^{(i,j)}$, 
number of dijets $N_{jj}^{W20}$ with $| m_{2j} - m_W| \le 20$ GeV, 
max dijet $p_\text{T}$ vector sum, 
max $\Delta R$ of any dijet pair,
3-jet invariant mass $m_{3j}^{(i,j,k)}$, 
jet multiplicity $n_{j}$,
jet $b$-tag multiplicity and
invariant mass $m_{2j}^{\Delta R^\text{min}}$ of $\Delta R^\text{min}$ dijet, with $i, j,k=\overline{1,6}$ corresponding to jets ordered according to the highest $p_{\rm T}$.
To ensure consistency in the number of variables across events, as well as to leverage  the power of the ML algorithm to find the best combination of variables, we consider all combinations for $m_{2j}$, $m_{3j}$ and $\Delta R$.

The classification of signal and background events is based on the trained ML model, which assigns to each event a probability score $( P )$ with values ranging from 0 to 1.
This score represents the likelihood that an event belongs to the signal category. To convert these continuous probability scores into discrete classifications, a series of thresholds or discriminator values $(D)$, are applied. Events with \( P > D \) are classified as signal, whereas those with \( P \leq D \) are classified as background. 

For this analysis, a wide range of discriminator values is considered, starting from 0.2, all the way to 0.99. These values allow for an evaluation of the model performance across different working points, with lower thresholds ({\it e.g.}, \( D = 0.80 \)) classifying a larger number of events as signal, leading to higher signal efficiency but lower purity. In contrast, higher thresholds ({\it e.g.},  \( D = 0.99 \)) result in a highly pure signal selection but at the cost of reduced efficiency. The signal and background 
distributions in the $0.20-0.80$ discriminator range are uniform, so we have decided to present the results for $D\geq0.80$. In order to maintain a good signal purity with a reasonable selection efficiency, we have set the higher boundary of the discriminant to $D=0.97$.

To estimate the expected number of signal and background events, each event is weighted by its process-specific cross section, the integrated luminosity, and the ratio of events passing the discriminator cut. We use the notation $\sev$ 
for the number of expected signal events, while the background counts, denoted by \( \bev \), correspond to events that the model classifies as signal, regardless of their true origin (misclassified background events).

\begin{ruledtabular} 
\begin{table}[ht]
\vspace*{-0.1mm}
\caption{\label{tab:ATLAS_BDT_RF}%
Number of events for signal ($\sev$) and background ($\bev$) obtained with the BDT and RF algorithms for the ATLAS detector parametrization at $\sqrt{s} = 13.6$ TeV and $\mathcal{L}=3000 \ \text{fb}^{-1}$, for $M_{S}=7.5$ TeV, $y_{uu}  B(\chi \to W b) = 0.1$, 
and $\widehat{m}_{min}=6$ TeV.
}
\centering
\begin{tabular}{llccccc}
& & $D$=0.80 & $D$=0.90 & $D$=0.95 & $D$=0.96 & $D$=0.97 \\[0.5mm] \hline
\\[-2mm]
\multirow{2}{2em}{BDT}
 & $\sev$ & 18.4 & 18.3 & 17.5 & 16.5 & 14.3 \\[0.5mm]
 & $\bev$ & 133 & 57.4 & 11.2 & 7.8 & 2.3 \\[0.5mm]
\hline
\\[-2mm]
\multirow{2}{2em}{RF} & $\sev$ & 18.3 & 18.3 & 17.7 & 16.9 & 14.7 \\[0.5mm]
 & $\bev$ & 149 & 73.1 & 17.2 & 10.4 & 5.2 \\[0.5mm]
\end{tabular}
\end{table}
\end{ruledtabular}

We simulated $10^5$ events for each process, signal or background, of which 80\% were used in training the models and the remaining 20\% of the total sample was used for performance evaluation. In the following, we are showing the $\sev$ and $\bev$ values obtained for various cases.
In Table~\ref{tab:ATLAS_BDT_RF} we present the number of events for signal and background obtained with BDT and RF for $D\geq0.80$.
Detector simulation is done here with the ATLAS parametrization of \textsc{Delphes} \cite{deFavereau:2013fsa},  
assuming $\sqrt{s} = 13.6$ TeV, $\mathcal{L}=3000 \ \text{fb}^{-1}$, 
$\ms = 7.5$ TeV, and $y_{uu}  B(\chi \to W b) = 0.1$. 
The other model parameters are fixed as in Eq.~(\ref{eq:benchmark}), and the \textsc{Pythia} phase-space cut is set here at $\widehat{m}_{min}=6$ TeV. 

From the results presented in Table \ref{tab:ATLAS_BDT_RF} we can conclude that both ML classification models, RF and BDT, generally provide us with comparable results for the discriminator $D$ in the interval $[0.80-0.97]$.
However, depending on the random process of model parameter initialization, the BDT could end up with slightly smaller errors in some runs. 
Even so, we choose to continue our study with RF because it is less prone to over-fitting \cite{hastie2009elements}.

\vskip -5mm
\begin{ruledtabular} 
\begin{table}[ht]
\caption{\label{tab:CMS_RF}%
$\sev$ and $\bev$ obtained with RF for the CMS detector parametrization of \textsc{Delphes}. The parameters are fixed as specified in the caption of Table \ref{tab:ATLAS_BDT_RF}.
}
\centering
\begin{tabular}{lccccc}
& $D$=0.80 & $D$=0.90 & $D$=0.95 & $D$=0.96 & $D$=0.97 \\[0.5mm] \hline
\\[-2mm]
$\sev$ & 18.3 & 18.3 & 17.7 & 16.9 & 14.7 \\[0.5mm]
$\bev$ & 122.1 & 52.6 & 14.6 & 10.8 & 1.6 \\ [0mm] 
\end{tabular}
\end{table}
\end{ruledtabular}

Table \ref{tab:CMS_RF} shows $\sev$ and $\bev$ values obtained with the RF algorithm for the CMS parametrization of the \textsc{Delphes} detector simulation. 
Because we obtain similar signal-to-background discrimination for the CMS and ATLAS detector parametrizations (compare Tables \ref{tab:CMS_RF} and  \ref{tab:ATLAS_BDT_RF}), in what follows we perform the signal selection efficiency study only for the ATLAS parametrization.

\section{\label{results}Signal selection efficiency}
\label{sec:effic}

In this section we study how the 
number of signal and background events ($\sev$ and $\bev$)  depend  on the RF and sample parameters. We first discuss the impact of variable weights on the ML discriminator $D$. Then, in Sec. 
\ref{subsection:center_of_mass} we compare $\sev$ and $\bev$ at two center-of-mass energies, $\sqrt{s}=13.6$ TeV and $14$ TeV, and investigate the relevance of the number of simulated events.
In \ref{subsection:jets}, we explore the $\sev$ and $\bev$  dependence on the number of jets used by the ML algorithm, and 
we outline the influence of the phase-space cuts. 
All the studies of this section are performed for $y_{uu} B(\chi \to W b) = 0.1$, with other parameters fixed as in Eq.~(\ref{eq:benchmark}). 

We use the k-fold cross-validation method to compute the errors of our ML algorithm. We work with 5 folds, splitting $10^5$ events data sample into $80\%$ for training and $20\%$ for testing.
In the case of $\sev$ we display only the mean values, but one should keep in mind that there are sizable uncertainties, especially from NLO effects and from PDFs at the large parton  momentum fractions associated with an ultraheavy resonance. The errors on $\sev$ due to the ML algorithm are below 1\%, so they are negligible compared to the uncertainties mentioned above. 

For $\bev$, in addition to the mean values, we display the errors due to the RF model performance as the standard deviation of all runs.
We emphasize that there are other large uncertainties on $\bev$, for example from QCD processes (see  Sec. \ref{sec:particles}) not included in the simulation.

\subsection{\label{feature importance}Feature importance}

To gain insight into the decision-making process of the Random Forest classifier, we examine the feature importance scores assigned to each input variable. The importance of a feature is determined using the Mean Decrease in Impurity (MDI) method \cite{breiman1984classification}, which quantifies how much each feature contributes to improving the splits in the RF decision trees. Features that are more frequently used for splitting the data and lead to better separation between signal and background are assigned higher importance values. The final feature importance scores are averaged over all the decision trees in the RF and normalized to sum to one.

We run tests on different sample folds, center-of-mass energies, values of diquark mass $\ms$, values of the phase-space cut $\widehat{m}_{\min}$, number $n_{j}$ of jets in the final state, and number $N_{\rm ev}$ of simulated events per process.
Although distinct runs lead to distinguishable feature lists, they consist mainly of the same variables. 
Only slight changes between their ranking are observed. 
We have found that the transverse momentum of the fourth and the third final-state jets $p_\text{T}^{(4)}$ and $p_\text{T}^{(3)}$ are always the first two with importance score in the $12\%-14\%$ range. 

This is not surprising given that the background only rarely includes four jets of high $p_\text{T}$, while the signal always includes four high-$p_\text{T}$ jets. Note that the fifth jet in the signal is often soft because the two jets arising from one $W$ decay in the signal are typically merged. Our parton-level simulation indicates a merging rate of about 96\% when the quark-jet matching is done for $\Delta R < 0.4$. Even at the reconstruction level, the effect remains pronounced: in roughly 64\% of the events, no more than five jets are observed, even in the presence of initial and final state radiation; see also Sec. IV C. The $W$-merged jets tend to have higher $p_{\rm T}$ than the $b$-jets, such that the third highest jet $p_{T}$ is almost twice as often a $b$-jet as it is a $W$-merged  jet. 

Other features with high importance score include: $m^{(1,i)}_{2j}$, $m^{\Delta R^{\text{min}}}_{2j}$, $p^{(2)}_{\text{T}}$, and $m^{(1,2,k)}_{3j}$, with $i\geq2$ and $k=3,4$. 
The importance score of each of these kinematic variables is below $8\%$.
Although we are using 75 input variables to take into account as many signal and background characteristics as possible, not all variables  have a significant impact on the decisions of the RF model.  Among the low-impact ones are variables related to the lower-$p_T$ jets (with $i \ge 5$), including $\Delta R^{(i,j)}$, $m^{(i,j)}_{2j}$, $m^{(i,j,k)}_{3j}$, $\eta^{(i)}$ and $\phi^{(i)}$.

\subsection{\label{subsection:center_of_mass}Center-of-mass energy and number of events}

The increase in the $\sigma_{\! _S}(\chi\chi)$ cross section with $\sqrt s$  (as shown in Fig. \ref{fig:cross_section_yuuB}) translates into higher values of the number of signal events $\sev$. At the same time, higher $\sqrt s$ leads to a larger number of background events $\bev$. Nevertheless, the six jets associated with the leading-order signal considered here have large transverse momenta, 
due to the very large mass of the $s$-channel resonance, so that the $\sev/\sqrt{\bev}$ ratio grows when $\sqrt s$ is increased above 13.6 TeV.

A comparison of the number of signal and background events  for $\sqrt s = 13.6$ TeV versus $\sqrt s = 14$ TeV is presented in Table \ref{tab:center-of-mass_dependency}, assuming an integrated luminosity $\mathcal{L}=3000 \ \text{fb}^{-1}$ 
in both cases.
While the center-of-mass energy planned for the HL-LHC is 14 TeV, the current Run 3 of the LHC is at $\sqrt{s}=13.6$ TeV. To be conservative, we will set $\sqrt{s}=13.6$ TeV for our feasibility study. However, it should be emphasized that even a slightly larger center-of-mass energy ({\it e.g.}, 14 TeV) would increase the HL-LHC sensitivity to ultraheavy particles such as the diquark studied here.

\begin{ruledtabular}
\begin{table}[t]
\caption{\label{tab:center-of-mass_dependency}%
Number of events for signal ($\sev$) and background ($\bev$) obtained for two center-of-mass energies, 13.6 TeV and 14 TeV, with $3000 \ \text{fb}^{-1}$ of simulated data, when 
$\ms = 7.5$ TeV, $\widehat{m}_{min}=6$ TeV, and $y_{uu} B(\chi \to W b) = 0.1$. Other model parameters are fixed as in Eq.~(\ref{eq:benchmark}).  \\[-2.5mm]
}
\centering
\footnotesize
\begin{tabular}{lccccc}
& $D=0.80$ & $D=0.90$ & $D=0.95$ & $D=0.96$ & $D=0.97$ \\[1mm]   
\hline
\\[-2.0mm]
\multicolumn{6}{c}{$\sqrt{s} = 13.6$ TeV} \\[0.5mm]
\hline\\[-2.5mm]
$\sev$ & 18.3 & 18.3 & 17.7 & 16.9  & 14.7 \\[0.5mm]
$\bev$ & 149{\tiny$\pm$}21 & 73.1{\tiny$\pm$}11.4 & 17.2{\tiny$\pm$}4.0 & 10.4{\tiny$\pm$}0.6 & 5.2{\tiny$\pm$}2.9 \\[0.5mm]
\hline\\[-2.0mm]
 \multicolumn{6}{c}{$\sqrt{s} = 14$ TeV} \\[0.5mm]
\hline\\[-2.5mm]
$\sev$ & 25.3 & 25.3 & 24.4 & 23.3 & 20.3 \\[0.5mm]
$\bev$ & 209{\tiny$\pm$}9 & 113{\tiny$\pm$}5 & 21.8{\tiny$\pm$}7.2 & 14.7{\tiny$\pm$}6.8 & 3.1{\tiny$\pm$}2.4   
\end{tabular}
\end{table}
\end{ruledtabular}

\begin{ruledtabular}
\begin{table}[h]
\caption{\label{tab:number_events}%
$\sev$ and $\bev$ dependence on the number of simulated events per process at $\sqrt{s} = 13.6$ TeV and $\mathcal{L}=3000 \ \text{fb}^{-1}$, for $\ms =7.5$ TeV and 
other parameters as in Table \ref{tab:center-of-mass_dependency}. The uncertainties for $\bev$ displayed here are only due to the errors of the ML algorithm (same type of errors for $\sev$ are below 1\%). \\[-2.5mm]
}
\centering
\footnotesize
\begin{tabular}{lccccc}
& $D=0.80$ & $D=0.90$ & $D=0.95$ & $D=0.96$ & $D=0.97$ \\[1mm]  
\hline
\\[-2.0mm]
\multicolumn{6}{c}{$N_{\rm ev}=10^5$} \\[0.5mm]
\hline \\[-2.0mm]
$\sev$ & 18.3 & 18.3 & 17.7 & 16.9 & 14.7 \\[0.5mm]
$\bev$ & 149{\tiny$\pm$}21 & 73.1{\tiny$\pm$}11.4 & 17.2{\tiny$\pm$}4.0 & 10.4{\tiny$\pm$}0.6 & 5.2{\tiny$\pm$}2.9 \\[0.5mm]
\hline \\[-2.0mm]
\multicolumn{6}{c}{$N_{\rm ev}=2 \times 10^5$} \\[0.5mm]
\hline \\[-2.0mm]
$\sev$ & 18.3 & 18.3 & 17.7 & 17.0 & 14.8 \\[0.5mm]
$\bev$ & 139{\tiny$\pm$}9 & 71.0{\tiny$\pm$}8.5 & 17.5{\tiny$\pm$}3.8 & 8.2{\tiny$\pm$}3.7 & 3.1{\tiny$\pm$}1.4 \\[0.5mm]
\hline \\[-2.0mm]
\multicolumn{6}{c}{$N_{\rm ev}=3 \times 10^5$} \\[0.5mm]
\hline\\[-2.0mm]
$\sev$ & 18.3 & 18.3 & 17.8 & 17.0 & 14.9 \\[0.5mm]
$\bev$ & 133{\tiny$\pm$}11 & 71.9{\tiny$\pm$}6.3 & 17.1{\tiny$\pm$}3.9 & 7.9{\tiny$\pm$}2.8 & 2.7{\tiny$\pm$}1.4 \\[0.5mm]
\end{tabular}
\end{table}
\end{ruledtabular}
Next, we analyze how the signal and background vary with  the size of the simulated sample.
In Table \ref{tab:number_events}
we show $\sev$ and $\bev$ obtained for different number $N_{\rm ev}$ of simulated events.
When the number $N_{\rm ev}$ of MC simulated events per process varies from $N_{\rm ev} = 10^5$ to $N_{\rm ev} = 3 \times 10^5$ and $D\leq 0.97$, we see that the number of background events is getting smaller for a fixed discriminator ($D$) value.
However, for a fixed value of  $D$, the $\sev$ variation between samples is almost negligible. 

Also shown in Table \ref{tab:number_events}
are uncertainties in the background due to the ML separation method. The dependence of these uncertainties on $N_{\rm ev}$, when it varies in the range $10^5$ to $3 \times 10^5$, is somewhat unstable due the 
sensitivity of the 
k-fold cross-validation method to RF decisions. 
 Using larger samples would be beneficial, but computationally too expensive.
Nevertheless, compared to the case $N_{\rm ev} = 5 \times 10^4$ (not shown in Table \ref{tab:number_events}), the background uncertainties are substantially smaller at $N_{\rm ev}=10^5$.
Although the signal-to-background selection is slightly improved for $N_{\rm ev} = 3 \times 10^5$ compared to $N_{\rm ev} = 10^5$, we have decided to continue our study with $N_{\rm ev} = 10^5$ samples in order to gain computational efficiency.

\subsection{\label{subsection:jets}Number of jets and phase-space cut}

The final state arising from the cascade decay of the $S_{uu}$ scalar into $(W^+b)(W^+b)$, with hadronic $W^+$ decays, consists in two pairs of nearly collinear quark jets and two $b$ jets.
Even though this signal process involves a 6-jet final state,
additional jets are produced in the same event due to multiparton interactions, and parton showering.
Notice that initial- and 
final-state radiation also contribute additional jets, but these are NLO effects, and are  not taken into account in this analysis.

In some events, the number of jets may be smaller due to possible collimation between some jets. In particular, each of the two $W^+$ bosons from the final state is typically boosted due to the large $S_{uu}$ mass. Thus, each $W^+$ often produces a single jet with a two-prong substructure.

\begin{ruledtabular}
\begin{table}[t]
\caption{\label{tab:errors_jets}
$S_{ev}$ and $B_{ev}$ for a few choices of the number of final state jets taken into account by the ML algorithm at $\sqrt{s} = 13.6$ TeV and 
$\mathcal{L}=3000 \ \text{fb}^{-1}$, for  $M_{S}=7.5$ TeV and $\widehat{m}_{min} = 7$ TeV. 
}
\centering
\footnotesize
\begin{tabular}{lccccc}
& $D=0.80$ & $D=0.90$ & $D=0.95$ & $D=0.96$ & $D=0.97$ \\[1mm]   
\hline
\\[-2.0mm]
\multicolumn{6}{c}{$n_j = 4$} \\[0.5mm]
\hline \\[-2.0mm]
$\sev$ & 18.3 & 18.3 & 18.0 &  17.6 & 16.0 \\[0.5mm]
$\bev$ & 11.9{\tiny$\pm$}0.3 & 11.9{\tiny$\pm$}0.3 & 1.9{\tiny$\pm$}0.7 & 1.0{\tiny$\pm$}0.4 & 0.34{\tiny$\pm$}0.20 \\[0.5mm]
\hline \\[-2.0mm]
\multicolumn{6}{c}{$n_j = 6$} \\[0.5mm]
\hline \\[-2.0mm]
$\sev$ & 18.3 & 18.3 & 18.0 & 17.6 & 16.0 \\[0.5mm]
$\bev$ & 15.0{\tiny$\pm$}1.3 & 8.0{\tiny$\pm$}4.6 & 2.03{\tiny$\pm$}0.80 & 1.1{\tiny$\pm$}0.5 & 0.18{\tiny$\pm$}0.05 \\[0.5mm]
\hline \\[-2.0mm]
\multicolumn{6}{c}{$n_j = 8$} \\[0.5mm]
\hline \\[-2.0mm]
$\sev$ & 18.3 & 18.3 & 18.1 & 17.6 & 16.1 \\[0.5mm]
$\bev$ & 18.9{\tiny$\pm$}1.7 & 6.46{\tiny$\pm$}1.49 & 2.08{\tiny$\pm$}0.62 & 1.2{\tiny$\pm$}0.6 & 0.17{\tiny$\pm$}0.07
\end{tabular}
\end{table}
\end{ruledtabular}

After the jets are ordered according to their transverse momentum, the ML algorithm takes into account a number $n_j$
of jets with highest $p_T$.
As shown in Table \ref{tab:errors_jets}, the number of signal events ($\sev$)
has only a very weak dependence on $n_j$, which is negligible within errors.
For $D\leq0.95$, we observe a better signal versus background discrimination for the $n_j=4$ case than for the other ones. When increasing the number of jets to $n_j=6$ and $n_j=8$, signal-to-background selection becomes comparable. Due to this finding we have decided to continue the current study with the first six most energetic jets taken into account by the ML algorithm.

\begin{figure}[t]
\centering
\includegraphics[width=0.47\textwidth]{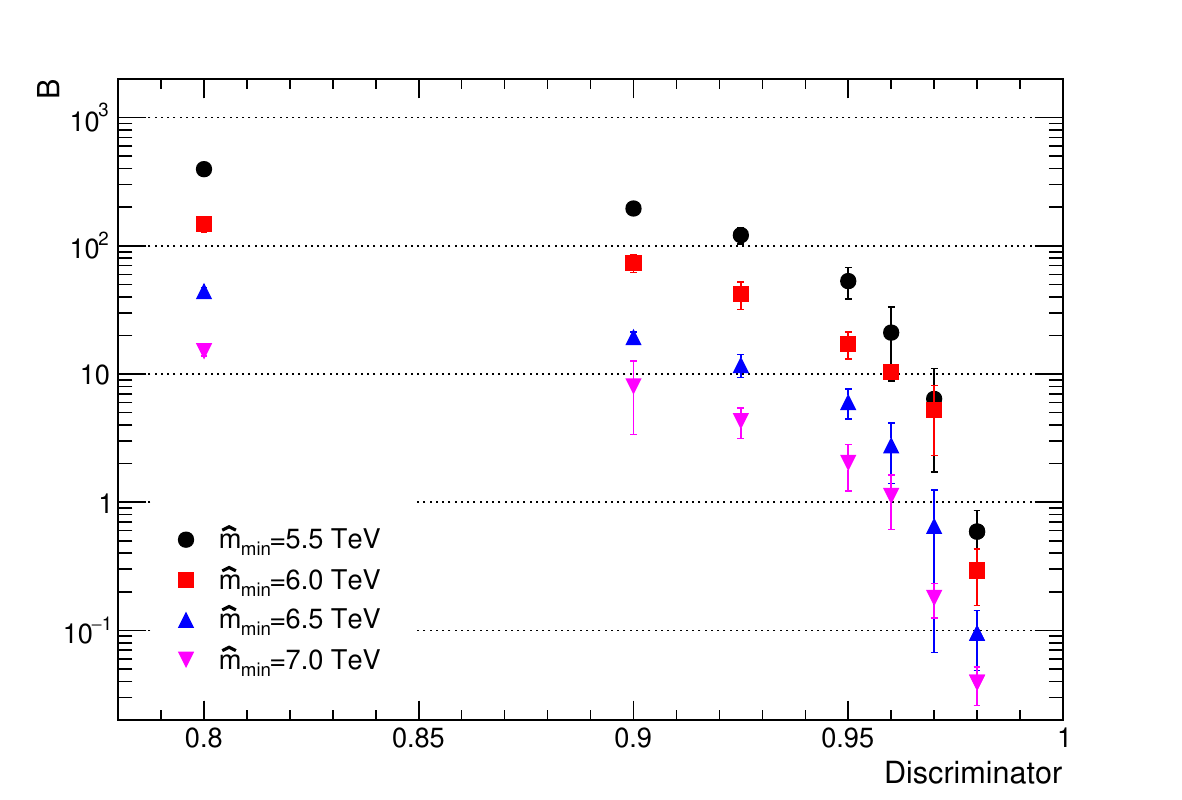}
    \caption{{\label{fig:phase_space}
     Number of background events ($\bev$) as a function of discriminator $D$, for different values of the 
     phase-space cut $(\widehat{m}_{min})$.
     The integrated luminosity is taken to be $\mathcal{L}=3000 \ \text{fb}^{-1}$ at $\sqrt{s} = 13.6$ TeV
     and the diquark mass is fixed at 
     $\ms = 7.5$ TeV.}}
\end{figure}

\begin{ruledtabular}
\begin{table}[ht]
\caption{\label{tab:errors_kfold}
$\sev$ and $\bev$ for different choices of the \textsc{Pythia} phase-space cut $\widehat{m}_{min}$  for $\ms = 7.5$ TeV and other parameters fixed as in Table \ref{tab:number_events}.
}
\centering
\footnotesize
\begin{tabular}{lccccc}
& $D=0.80$ & $D=0.90$ & $D=0.95$ & $D=0.96$ & $D=0.97$ \\[1mm]   
\hline
\\[-2.0mm]
\multicolumn{6}{c}{$\widehat{m}_{min}=$ 5.5 TeV} \\[0.5mm]
\hline \\[-2.0mm]
$\sev$ & 18.3 & 18.3 & 17.8 & 17.0 & 14.8 \\[0.5mm]
$\bev$ & 397{\tiny$\pm$}46 & 196{\tiny$\pm$}19 & 53.1{\tiny$\pm$}14.8 & 21.0{\tiny$\pm$}12.2 & 6.4{\tiny$\pm$}4.7 \\[0.5mm]
\hline \\[-2.0mm]
\multicolumn{6}{c}{$\widehat{m}_{min}=$ 6 TeV} \\[0.5mm]
\hline \\[-2.0mm]
$\sev$ & 18.3 & 18.3 & 17.7 & 16.9 & 14.7  \\[0.5mm]
$\bev$ & 149{\tiny$\pm$}21 & 73.1{\tiny$\pm$}11.4 & 17.2{\tiny$\pm$}4.0 & 10.4{\tiny$\pm$}0.6 & 5.2{\tiny$\pm$}2.9 \\[0.5mm]
\hline \\[-2.0mm]
\multicolumn{6}{c}{$\widehat{m}_{min}=$ 6.5 TeV} \\[0.5mm]
\hline \\[-2.0mm]
$\sev$ & 18.3 & 18.3 & 17.8 & 17.1 & 15.1 \\[0.5mm]
$\bev$ & 44.4{\tiny$\pm$}2.9 & 19.4{\tiny$\pm$}1.8 & 6.1{\tiny$\pm$}1.6 & 2.8{\tiny$\pm$}1.4 & 0.65{\tiny$\pm$}0.59 \\[0.5mm]
\hline \\[-2.0mm]
\multicolumn{6}{c}{$\widehat{m}_{min}=$ 7 TeV} \\[0.5mm]
\hline\\[-2.0mm]
$\sev$ & 18.3 & 18.3 & 18.0 & 17.6 & 16.0 \\[0.5mm]
$\bev$ & 15.0{\tiny$\pm$}1.3 & 8.0{\tiny$\pm$}4.6 & 2.0{\tiny$\pm$}0.8 & 1.1{\tiny$\pm$}0.5 & 0.18{\tiny$\pm$}0.05 
\end{tabular}
\smallskip
\end{table}
\end{ruledtabular}

The background simulation employed here strongly depends on the  phase-space cut $\widehat{m}_{min}$. 
In Fig. \ref{fig:phase_space} we show that $\bev$ values (at $\mathcal{L}=3000 \ \text{fb}^{-1}$)
decrease for higher phase-space cuts $\widehat{m}_{min}$, which means better performance of the ML algorithm.
Table \ref{tab:errors_kfold} shows the $S_{ev}$ and $B_{ev}$ values for a fixed diquark scalar mass and different $\widehat{m}_{min}$. The highest number of signal events and the lowest number of background events are obtained for the highest phase-space cut. 
The uncertainties in the background shown in Table \ref{tab:errors_kfold} are only those due to the ML algorithm. Their dependence 
 on $\widehat{m}_{min}$ is again somewhat unstable (see Sec.~\ref{subsection:center_of_mass}).

\section{\label{sec:obs_potential} $S_{uu}$ Observation potential}

The purpose of this study is to investigate the discovery potential of a heavy diquark scalar $S_{uu}$ which decays into a pair of vectorlike quarks, as described by the model presented in \cite{Dobrescu:2019nys}. CMS Collaboration \cite{CMS:2022usq}  observed two events with four hadronic jets that form a very large invariant mass $M_{4j}$, around 8 TeV, and are paired in two dijets, each having an invariant mass $M_{2j}$ near 2 TeV.
While the ATLAS search \cite{ATLAS:2023ssk} in the same final state yielded no events with $M_{4j}$ at 8 TeV, it reported an intriguing event with $M_{4j} \approx 6.6$ TeV that appears consistent with a pair-produced vectorlike quark of mass near 2 TeV (for related interpretations, see the CMS results in  \cite{CMS:2025hpa}, and the $S2\chi A$ model in \cite{Dobrescu:2024mdl}). 

This motivates our choice of the  masses for the new particles, 
$M_S$ in the $7-8.5$ TeV range, and $m_\chi \approx 2$ TeV. Even if the events mentioned above will turn out to be just a very unlikely fluctuation of the QCD background, the search for resonantly produced vectorlike quarks discussed in this paper  is a well-motivated exploration of physics at scales in the multi-TeV range.

We present here our results regarding the observation potential of the ultraheavy diquark scalar $S_{uu}$,  
obtained with 3000 fb$^{-1}$ of simulated data at $\sqrt{s}=13.6$ TeV based on a sliding phase-space cut given by $\widehat{m}_{min}=\ms-0.5$ TeV. 
The number of signal and background events shown in Table \ref{tab:golden_case} corresponds to different values of $\ms$. The signal is shown there for two values of the dimensionless parameter
$y_{uu}B(\chi\to Wb)$: $0.1$ and $0.2$, with the number of signal events labeled by $\sev(0.1)$ and $\sev(0.2)$, respectively.
Because this observation potential study relies on an incomplete  simulation of the background, and uses the \textsc{Delphes} detector simulation \cite{deFavereau:2013fsa}, which has larger uncertainties compared to full detector simulations, the k-fold cross-validation errors of the RF are not included in Table \ref{tab:golden_case}. 

\begin{ruledtabular}
\begin{table}[b]
\caption{\label{tab:golden_case}
Number of signal events $\sev(0.1)$ and $\sev(0.2)$, for $y_{uu}B(\chi\to Wb) = 0.1$ and $0.2$, respectively, number of background events $\bev$, predicted for different values of the diquark mass, with 3000 fb$^{-1}$ of data at $\sqrt{s}=13.6$ TeV, and a phase-space cut  $\widehat{m}_{min}=\ms-0.5$ TeV. \\[-2.4mm]
}
\centering
\footnotesize
\begin{tabular}{cccccc}
& \hspace*{-0.4cm} $ D=0.80 \!\!$ & $\!\! D=0.90\!$ & $\!\! D=0.95\!$ & $\!\! D=0.96\!$ & $\!\! D=0.97 \!\!$ \\[1mm]   
\hline
\\[-1.61mm]
\multicolumn{6}{c}{$M_{S}=7.0$ TeV} \\[1mm]
\hline\\[-2.6mm]
 $\sev(0.1)$ &  43.9 & 43.8 & 43.0 & 41.9  & 37.9   \\[1mm]
 $\sev(0.2)$ & 197 & 196 & 193 & 188 & 170  \\[1mm]
 $\bev$   &   42.4 & 17.9  & 5.05 & \ \ \ 3.96  & \ \ \ 0.79 \\[1mm]
\hline\\[-1.61mm]
 \multicolumn{6}{c}{$M_{S}=7.5$ TeV} \\[1mm]
\hline\\[-2.6mm]
 $\sev(0.1)$ & 18.3 & 18.3 & 18.0 & 17.6 & 16.0 \\[1mm]
 $\sev(0.2)$ & 81.9 & 81.8 & 80.5 & 78.6 & 71.5 \\[1mm]
 $\bev$ & 15.0 & 8.00 & 2.03 & 1.12 & 0.18 \\[1mm]
\hline\\[-1.61mm]
 \multicolumn{6}{c}{$M_{S}=8.0$ TeV} \\[1mm]
\hline\\[-2.6mm]
 $\sev(0.1)$ & 7.20 & 7.20 & 7.10 & 6.95 & 6.35 \\[1mm]
 $\sev(0.2)$ & 32.4 & 32.4 & 31.9 & 31.3 & 28.5 \\[1mm]
 $\bev$ & 5.34 & 5.34 & 0.67 & 0.36 & 0.09 \\[1mm]
\hline\\[-1.61mm]
\multicolumn{6}{c}{$M_{S}=8.5$ TeV} \\[1mm]
\hline\\[-2.6mm]
$\sev(0.1)$ & 2.63 & 2.63 & 2.59 & 2.54 & 2.33 \\[1mm]
$\sev(0.2)$ & 11.8 & 11.8 & 11.6 & 11.4 & 10.5 \\[1mm]
$\bev$ & 1.72 & 1.72 & 0.17 & 0.09 & 0.02 \\[1mm]
\end{tabular}
\end{table}
\end{ruledtabular}

For $y_{uu}B(\chi \rightarrow Wb)=0.1$ and $M_S = 8$ TeV the 95\% upper limit of our background estimate is about 0.4 events when the discriminator is $D =0.96$.
At this point in parameter space of the model with $S_{uu}$ and $\chi$, the expected number of events is  $\sev(0.1) + \bev \approx 7.3$.
The probability for the SM background to have a large upward fluctuation such that 7 or more events are observed, in the absence of a new physics signal, can be computed using Poisson statistics (for an analysis of potential complications, see  \cite{Feldman:1997qc}) and is given by $1.1 \times 10^{-7}$. This is slightly above a 5$\sigma$ excess, and can be considered a discovery.

The large uncertainties in the background may degrade the significance of the excess, but  when the discriminator is increased to $D =0.97$ the background decreases by a factor of four while the signal decreases only by about 10\%. Even if $\bev$ were 
0.25 events (which is almost three times larger than the estimated background at $D =0.97$ and $\ms = 8$ TeV),  
its fluctuation into 6 or more observed events would have a probability of only $2.7 \times 10^{-7}$. In that case the new physics model discussed here, which predicts $\sev(0.1) + \bev \approx 6.6$, would be again favored over the SM at the 5$\sigma$  level.

Based on these results, we conclude that our ML method 
provides a good HL-LHC discovery potential for an $S_{uu}$ of mass near 8 TeV, which decays into $(W^+b)(W^+b)$ with a final state consisting of six jets, provided $y_{uu}B(\chi \rightarrow Wb) \ge 0.1$. Given that the typical branching fraction of a vectorlike quark that mixes with the top quark is $B(\chi \rightarrow Wb) \approx 50\%$, the signal discussed here can be discovered even when $y_{uu}$ is as small as 0.2.

For a heavier $S_{uu}$, with mass near 8.5 TeV, a larger $y_{uu}$ may be necessary for discovery. As can be seen from 
Table \ref{tab:golden_case}, 
a factor of 2 increase in $y_{uu}$ is already sufficient for discovery at $M_S = 8.5$ TeV, as it would lead to about 11 signal events, while the background is below 0.2 events for an ML discriminator $D \ge 0.95$. 

The benchmark choices in Eq.~\ref{eq:benchmark_sensitivity} are rather conservative, focusing on a $y_{uu}$ smaller than the central values indicated by the events with dijet pairs \cite{Dobrescu:2024mdl}. For this reason, as well as due to the total branching fraction of only 11\% for $\chi\chi \to (W^+b)\, (W^+b)  \to (jjb)\, (jjb)$,   
the full integrated luminosity of HL-LHC would be required for discovery in this 
final state. If additional events with dijet pairs near 8 TeV will be observed in Run 3, then probing  
our $(W^+b) \, (W^+b)$ channel would continue to be important, but for a larger $y_{uu}$ and a smaller $\chi \to W^+b$ branching fraction.

Existing ATLAS 
\cite{ATLAS:2024gyc} and CMS \cite{CMS:2022fck} searches for vectorlike quarks (which set lower mass limits on $\chi$ in the   $1.4-1.7$ TeV range; other limits are reviewed in \cite{Alves:2023ufm}), which include $(W^+ b)(W^- \bar b)$ final states, are not well-suited for setting  limits on the resonantly produced $\chi\chi$ signal discussed here because the ultraheavy $S_{uu}$ diquark forces the vectorlike quarks to be boosted, changing substantially the kinematic distributions of the final states. Likewise, existing ATLAS \cite{ATLAS:2024kqk} and CMS \cite{CMS:2024ldy}  searches for 6-jets final states do not take advantage of  the large $p_T$'s of the jets originating from  a cascade decay of an ultraheavy $s$-channel resonance. Thus, it is important that the experimental collaborations perform dedicated searches for the signal analyzed here. More generally, our results motivate  searches for ultraheavy resonances in multijet final states that include $b$-tagged jets, and jets with substructure.

We emphasize that in this work  we focused only on the fully-hadronic final state arising from the $pp\to S_{uu}\to \chi\chi \to (W^+b) \, (W^+b)$ process. Semileptonic and fully leptonic final states of the $W^+$ pair may be included to increase the sensitivity of the  searches in the $(W^+ b)(W^+ \bar b)$ final states. Furthermore, as mentioned in \cite{Dobrescu:2019nys}, the sensitivity to an ultraheavy $S_{uu}$ may be improved by taking into account other  decay modes of $\chi$, for example into $tZ$ and $th^0$.

\section{Conclusions}
\label{sec:conclusions}

We studied the discovery potential of LHC experiments for resonantly produced vectorlike quarks, in the case where the $s$-channel resonance is a diquark scalar particle, $S_{uu}$, with mass in the 
$7-8.5$ TeV range. Such an 
ultraheavy particle may be   produced with a sufficiently large cross section at the LHC because it has a coupling to two up quarks, and thus benefits from the largest PDF near the kinematic limit of the collider. 

Previous studies of resonantly produced vectorlike quarks \cite{Dobrescu:2018psr, Dobrescu:2024mdl}
considered their exotic decay into two jets, motivated by the CMS observation of two remarkable events with $M_{4j} \approx 8$ TeV. Here we analyzed the more standard decay of a charged-2/3 vectorlike quark into $W^+b$, which is often expected to have the largest branching fraction (based on the assumption that mass mixing with the top quark is sufficiently  large  \cite{Han:2003wu}).

Given that the $S_{uu}$ resonance can be reconstructed when both $W^+$ bosons decay hadronically, in this paper we focused on the 6-jet final state arising from the  $pp\to S_{uu}\to \chi\chi \to (W^+b) \, (W^+b)$ process.
For the signal selection study we used Machine Learning models trained to discriminate against background sources. 
In particular, we investigated 
the impact of the variables weight, centre-of-mass energy, MC sample size, final state jet multiplicity, phase space selection, and diquark scalar mass on the Random Forest Machine Learning algorithm performance.

Our results indicate that searches at ATLAS or CMS in the above 6-jet final state with a luminosity of $3000 \ \text{fb}^{-1}$ 
may discover or rule out  
a diquark scalar of mass near 8 TeV even when its coupling to up quarks is as small as $y_{uu} \approx 0.2$.
This is a promising result for the scientific impact of future searches at the HL-LHC in this channel. Furthermore, it is likely that the ATLAS and CMS experimental collaborations would obtain more precise results using their own simulation framework, and may use additional features of the proposed signal to improve the sensitivity of future searches for resonantly produced vectorlike quarks.

\section*{Acknowledgments}
We thank Ethan Cannaert, John Conway, Patrick Fox, Robert Harris, Max Knobbe, and Julien Maurer for valuable comments and suggestions.
The work of I.D., C.A., I.-M.D. and M.-S.F. was supported by IFIN-HH under Contract No.~PN-23210104 with the Romanian Ministry of Education and Research.
The work of B.D. was supported by 
Fermi Forward Discovery Group, LLC under Contract No.~89243024CSC000002 with the U.S. Department of Energy, Office of Science, Office of High Energy Physics.

\providecommand{\noopsort}[1]{}\providecommand{\singleletter}[1]{#1}%

\end{document}